\documentclass[12pt,nofootinbib]{revtex4}

\usepackage{amsfonts}    
\usepackage{amssymb}
\usepackage{latexsym}
\usepackage{graphicx}
\usepackage[usenames,dvipsnames]{color}
\usepackage[english]{babel}
\newcommand{\al}{\alpha}

\newcommand{\ep}{\epsilon}

\newcommand{\ga}{\gamma}

\newcommand{\De}{\Delta}

\newcommand{\rar}{\rightarrow}

\begin{document}

\title{Three-body quantum Coulomb problem: analytic continuation}
\author{A.V.~Turbiner}

\email{alexander.turbiner@stonybrook.edu, turbiner@nucleares.unam.mx}
%
%
\affiliation{Instituto de Ciencias Nucleares, Universidad Nacional
Aut\'onoma de M\'exico, Apartado Postal 70-543, 04510 M\'exico, D.F., Mexico}
\author{J.C.~Lopez Vieyra}
\email{vieyra@nucleares.unam.mx}
\affiliation{Instituto de Ciencias Nucleares, Universidad Nacional
Aut\'onoma de M\'exico, Apartado Postal 70-543, 04510 M\'exico, D.F., Mexico}

\author{H.~Olivares Pil\'on}
\email{horop@nucleares.unam.mx}
\affiliation{Departamento de F\'isica, Universidad Aut\'onoma Metropolitana-Iztapalapa,
Apartado Postal 55-534, 09340 M\'exico, D.F., Mexico}

\begin{abstract}
\vskip 1.cm
  The second (unphysical) critical charge in the 3-body quantum Coulomb system of a nucleus of positive charge $Z$ and mass $m_p$, and two electrons,
  predicted by F~Stillinger has been calculated to be equal to $Z_{B}^{\infty}\ =\ 0.904854$ and $Z_{B}^{m_p}\ =\ 0.905138$ for infinite and finite (proton) mass $m_p$, respectively. It is shown that in both cases, the ground state energy $E(Z)$ (analytically continued beyond the first critical charge $Z_c$, for which the ionization energy vanishes, to $Re Z < Z_c$) has a square-root branch point with exponent 3/2 at $Z=Z_B$ in the complex $Z$-plane. Based on analytic continuation, the second, excited, spin-singlet bound state of negative hydrogen ion H${}^-$ is predicted to be at -0.51554 a.u. (-0.51531 a.u. for the finite proton mass $m_p$). The first critical charge $Z_c$ is found accurately for a finite proton mass $m_p$ in the Lagrange mesh method, $Z^{m_p}_{c}\ =\ 0.911\, 069\, 724\, 655$.
\end{abstract}

\maketitle

\newpage

We consider the 3-body quantum Coulomb system of two electrons $(e, m)$ and a (heavy) positive charge $Z$ of mass $m_p$, $(Z, m_p)$, where\ $m \ll m_p$. It is one of the most fundamental systems in theoretical physics. In atomic units $|e|=1$. Usually, this system is called the helium isoelectronic sequence, among many different names used. We denote this system as $(Z, e, e)$ and prefer to call it the {\it two-electron sequence}.
This system is fundamental: it includes the species like H${}^-$, He, Li${}^+$ etc, which play exceptionally important role in Nature. The non-relativistic Hamiltonian has the form
\begin{equation}
\label{H}
    {\cal H}\ =\ -\frac{1}{2} (\De_1 + \De_2)\ -\ \frac{\De}{2 m_p}  \ -\ \frac{Z}{r_{1}}\ -\
    \frac{Z}{r_{2}}  \ +\ \frac{1}{r_{12}}\ ,
\end{equation}
when written in atomic units with electronic mass $m=1$, where $r_1(r_2)$ is the distance from the charge
$Z$ to the first (second) electron, $r_{12}$ is interelectron distance, $\De_1 (\De_2)$ is the Laplacian
which describes the kinetic energy of the first (second) electron, and $\De$ is the Laplacian which describes
the kinetic energy of charge $Z$. The configuration space of (\ref{H}) is ${\bf R}^9$, but after separation
of centre-of-mass motion, we arrive at a six-dimensional configuration space of the relative motion.
Note that in the static limit $m_p \rar \infty$ the non-relativistic Hamiltonian (\ref{H}) degenerates to the form
\begin{equation}
\label{Hs}
    {\cal H}\ =\ -\frac{1}{2} (\De_1 + \De_2) \ -\ \frac{Z}{r_{1}}\ -\
    \frac{Z}{r_{2}}  \ +\ \frac{1}{r_{12}}\ .
\end{equation}
Due to charge quantization, the physical system occurs when $Z$ takes integer values, $Z=1,2,\ldots $.
At $Z \geq 1$ the energy of a bound state $E(Z)$ is a smooth function proportional to $(-Z^2)$ at large $Z$. Long ago it was given a rigorous mathematical proof in \cite{Kato:1980} that for $m_p=\infty$ the function $E(Z)/Z^2$ is the analytic function in vicinity of $1/Z=0$.
There is still unresolved challenge to find the radius of convergence in $1/Z$ expansion and establish nature of the singularity(ies) on its circle of convergence.
The lower edge of continuum (the threshold energy) is given by the ground state energy of the $Z$-hydrogen atom $(Z,e)$, $E_{th}= - Z^2 \frac{m_p}{2(1+m_p)}$~. It is well-known that there exists a certain charge $Z$ - called the critical charge $Z_c$ - for which the ground state energy of a two-electron sequence degenerates with the lower edge of continuum, thus, spontaneous ionization becomes possible. Quite recently, in the static limit (\ref{Hs}), the critical charge was accurately calculated (with triple basis sets containing up to 2276 terms, see \cite{Drake:2014}) and verified in independent calculation (using the Lagrange mesh method with lattice size $92 \times 92 \times 20$, see \cite{OT-PLA:2014}) within 12 significant figures
\begin{equation}
\label{zcrit}
      Z^{\infty}_{c}\ =\ 0.911\, 028\, 224\, 077\ .
\end{equation}
It was proved in \cite{Simon:1977} that at $Z=Z^{\infty}_{c}$ the Schr\"odinger equation has normalizable eigenfunction. In \cite{Drake:2014} this result was verified - it was explicitly stated that the trial function (triple basis set containing 2276 terms) is normalizable and localized, $<r_1> \sim 1/\al_1 \sim 1$\,a.u. and $<r_2> \sim 1/\al_2 \sim 5$\,a.u. , see below (6)-(7). It implies that the probability to meet electron at infinity is equal to zero, thus, the spontaneous  ionization is absent.

The critical charge $Z_c$ for the case of the finite mass $m_p$ was studied in \cite{Kais} for some particular values of $m/m_p$ and in \cite{Moini:2014} for $m_p \gtrsim 1$ and $m=1$. However, only recently, it was calculated accurately for the mass of proton,
$m_p = 1836.152 672 45 m_e$, where $m_e=1$ was set for the electron mass, using a 2856-term trial function (the convergence in energy was checked up to 11 s.d.) \cite{King:2015}
\begin{equation}
\label{zcrit-mp}
      Z^{m_p}_{c}\ =\ 0.911\, 069\, 7(3)\ ,
\end{equation}
which now we confirm (and improve) in the Lagrange mesh method \cite{Baye:2015} (see for details \cite{TLO:2016}) (the convergence in energy is checked up to 12 s.d. including, the maximal lattice size is
$92 \times 92 \times 20$)
\begin{equation}
\label{zcrit-mp-us}
      Z^{m_p}_{c}\ =\ 0.911\, 069\, 724\, 655\ ,
\end{equation}
Note that finite-mass effects are sufficiently small $\sim 10^{-5}$ changing the fifth figure in the critical charge, see (\ref{zcrit}) and (\ref{zcrit-mp-us}).

It is evident that when Z gets slightly smaller than $Z_c$ the bound (stationary) state becomes
quasi-stationary state (or Gamow state), characterized by complex energy, for discussion see
\cite{BZP,BZP2}. Its wavefunction is complex: it is mostly localized in the Coulomb well and it
corresponds to outgoing spherical wave at large $r_2 (r_1)$ at fixed $r_1 (r_2)$, $\sim  e^{-ikr_2}
\big(e^{-ikr_1}\big) $.  The imaginary part of energy is defined by tunneling rate from the Coulomb
well to infinity. In principle, it can be found via multidimensional WKB method, it should be
exponentially small, $\sim (Z_c -Z)^a e^{-\frac{b}{(Z_c -Z)^c}}$, where $b,c>0$ at $Z \rar Z_c-$. We
are not aware about any concrete theoretical calculations done so far. Hence, we have two different
spectral problems which can be summarized as follows: (i) for $Z > Z_c$ we have a problem of bound
states looking in the Schr\"odinger equation for solutions in the Hilbert space with (real) energies
$E=E_{>}(Z)$, (ii) for $Z < Z_c$ we have a
problem of quasi-stationary states looking in the Schr\"odinger equation for solutions corresponding to outgoing spherical wave with (complex) energies $E=E_{<}(Z)$. Thus, the boundary conditions for these two problems are completely different. In principle, there is no reason that these two analytic functions should coincide $E_{>}(Z)=E_{<}(Z)$. Furthermore, the function $E_{<}(Z)$ approaching to $Z \rar Z_c-$ is complex with exponentially small imaginary part (it indicates to the essential singularity at $Z = Z_c$), while the $E_{>}(Z)$ is real and approaching from the right to $Z \rar Z_c+$ behaves linearly \cite{Simon:1977}. It looks like a type of Kosterlitz-Thouless phase transition (of the infinite order) occurs at $Z=Z_c$:  the functions $E_{<}(Z)$ and $E_{>}(Z)$ and {\it all} their derivatives match at $Z=Z_c$.
The goal of our study is to find the function $E_{>}(Z)$ - the energy of the ground state - its analytic properties. We will not touch the function $E_{<}(Z)$ either theoretically, or numerically.

\begin{figure}
\begin{center}
 \includegraphics[width=3.2in,angle=0]{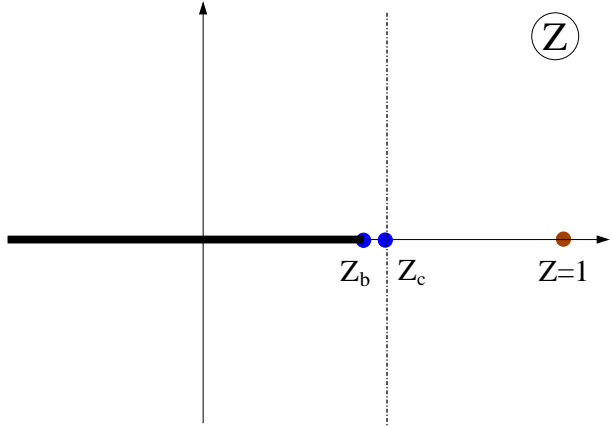}
\end{center}
 \caption{
 \label{E-ground-state}
 $Z$-complex plane of the ground state energy $E=E_{>}$, see text, the branch cut shown by bold wide
black line; vertical line passing through $Z_c$ indicates the line of discontinuity
(non-analyticity) between $E_{>}(Z)$ and  $E_{<}(Z)$ (see text).
}
\end{figure}

In remarkable papers of 1966 and 1974, F.H.~Stillinger and D.K.~Stillinger
\cite{Stillinger:1966,Stillinger:1974}, made an extended analysis of the ground state energy
obtained variationally as a function of continuous charge $Z$ and predicted the existence of the
second critical charge $Z_B$ - the charge for which the ground state eigenfunction looses its
normalizability, see \cite{Stillinger:1966}.
The trial function that was used for the analysis, was the celebrated Hylleraas-Eckart-Chandrasekhar function with a property of clusterization - in average one electron is closer to the charge center than the other, they are called innermost and outermost electrons,
\begin{equation}
\label{GS}
      \Psi_{trial} (r_1, r_2, r_{12})\ =\ \Psi_0 (r_1, r_2) + \Psi_0 (r_2, r_1)\ ,\ \Psi_0 (r_1, r_2)=e^{- \al_1 r_1 - \al_2 r_2}\ ,\ \al_1 \neq \al_2\ ,
\end{equation}
where $\al_{1,2}(Z)$ are two variational parameters, both real, found in the minimization of the energy functional for a fixed $Z$. The $Z$ dependence of the parameters $\al_{1,2}$ was quite non-trivial. It must be noted the important technical moment which made the analysis feasible: all integrals involved are evaluated analytically and are rational functions in $\al_{1,2}$. Hence, the variational energy is a simple analytic function of parameters $\al_{1,2}$  being the ratio of two polynomials in variables $\al_{1,2}$. It allows analytic considerations of the variational energy\footnote{ The same is correct for the function (\ref{GSB}), see below}.
Stillinger and Stillinger showed the existence of two distinct critical charges: (i) $Z_{c}(=0.9538)$ for which the ionization energy vanishes, hence, the ground state energy is equal to $-Z_{c}^2/2$ and coincides (degenerates) with the lower bound of continuum (the ground state energy of the $Z_c$-Hydrogen atom, $(Z_c,e)$), and (ii) $Z_{B}(=0.9276)$ for which the trial function (\ref{GS}) becomes non-normalizable: minimal $\al_{1}$ (or $\al_{2}$) vanishes, for illustration see Fig.~\ref{E-ground-state}. It was also demonstrated that the variational ground state energy $E_{var}=E(Z)$ is a regular function at $Z=Z_{c}$, while at $Z=Z_{B}$ it has a square-root branch point with exponent 3/2 with the branch cut going along real axis to the negative direction, $Z \in (-\infty, Z_{B}]$. Recently their analysis was extended: it was checked that, assuming the parameters $\al_{1,2}$ in (\ref{GS}) can be complex, the minimum of energy functional is always reached for real $\al_{1,2}$ for $Z>Z_B$.

The ground state variational energy can be constructed for a more general trial function with the electronic correlation explicitly included \cite{Bruno:2014},
\begin{equation}
\label{GSB}
  \Psi_0 (r_1, r_2, r_{12})\ =\ (1 + c r_{12}) e^{- \al_1 r_1 - \al_2 r_2 - \ga r_{12}}\ ,
            \ \al_1 \neq \al_2\ ,
\end{equation}
(c.f. (\ref{GS})), where $\al_{1,2}(Z), \ga(Z), c(Z)$ are four variational parameters.
Similarly to (\ref{GS}), the ground state was characterized by two distinct critical charges, $Z_{c} \neq Z_{B}$, where $Z_{c}(=0.9195)$ is regular point while $Z_{B}(=0.8684)$ is the square-root branch point with exponent 3/2. Thus, the introduction of two extra parameters $c, \ga$ to the trial function (\ref{GS}) shifts significantly the position of each critical point without changing the analytic properties. Note that both trial functions (\ref{GS}), (\ref{GSB}) lead to sufficiently accurate ground energy for $Z=1$ (which is a weakly-bound state): $-0.5133$ and $-0.5260$, respectively, while $E_{exact}=-0.52775$\,a.u. and for $Z=2$ : $-2.8757$ and $-2.9019$, respectively, while $E_{exact}=-2.9037$\,a.u. In fact, (\ref{GS}) and (\ref{GSB}) are the most accurate trial functions among two and four-parametric trial functions for any integer $Z \in [1,\infty)$: they give 1-2 and 2-3 correct decimal digits in ground state energy, respectively. Thus, both variational energies provide highly accurate, uniform(!)
approximation of the exact energy in $Z \in [1,\infty)$. Striking common property of both approximate variational energies as analytic functions of $Z$ is the absence of singularity at $Z=Z_c$. Note that the existence of two critical charges was also demonstrated in the Born-Oppenheimer approximation $m_p \rar 0$ (two-center, H$_2^+$-type case) \cite{TM:2012}.

The analysis of both variational energies $E_{var}(Z)$ obtained with (\ref{GS}), (\ref{GSB}) did not indicate the existence of other square-root branch points at finite $Z$ on the complex $Z$-plane \cite{Bruno:2014}.
Hence, one can draw the conclusion, in accordance with the Landau-Zener theory of the level
crossings (for discussion see e.g. \cite{BW,BW2}), that the point $Z_{B}$ seems to be a natural
candidate for the point of the level crossing of the two lowest spin-singlet, $1^1S$ and $2^1S$
states. However, the analytic continuation of the energy to the real positive axis of the second
sheet of the Riemann surface in $Z$ in the vicinity of $Z_{B}$ is not fully supported by the
analytic continuation of trial functions (\ref{GS}), (\ref{GSB}). Although these functions remain
normalizable after analytic continuation, they do not develop a nodal surface, thus, they are
NON-orthogonal to the ones corresponding to the ground state at given $Z$ and need to be
orthogonalized.

Extending the analysis of \cite{Stillinger:1966,Bruno:2014}, one can derive that for both cases (\ref{GS}) and (\ref{GSB}) the variational energy in $Z$ has two essentially different functional expansions near $Z_{c}$ and $Z_{B}$, namely:
\begin{equation}
\label{Zcr}
      E(Z) \ =\ -\frac{Z_{c}^2}{2} + a_1 (Z - Z_{c})+ a_2 (Z - Z_{c})^2 + a_3 (Z - Z_{c})^3 + \ldots\ ,
\end{equation}
which is the Taylor expansion: it indicates the absence of a singularity at $Z=Z_{c}$, and
\begin{equation}
\label{ZB}
    E(Z) \ =\ b_0 + b_1 (Z - Z_{B}) + c_1 (Z - Z_{B})^{\frac{3}{2}} + b_2 (Z - Z_{B})^2 +
    c_2 (Z - Z_{B})^{\frac{5}{2}} + \ldots\ ,
\end{equation}
for $Z \geq Z_{B}$, which is the so-called Puiseux expansion (the expansion in fractional degrees). The latter indicates the existence of the square-root branch point with exponent 3/2. In absence of other singularities in the vicinity of $Z_{c}$ other than at $Z=Z_B$, the radius of convergence of the expansion (\ref{Zcr}) has to be equal to $(Z_{c}-Z_B)$. Since the continuum starts at $E_{th} = -Z_{c}^2/2$, the presence of the bound state at $E \geq E_{th}$ indicates the phenomenon of the existence of a bound state {\it embedded} to continuum \cite{Stillinger:1975}
\footnote{Recently, this unexpected statement was supported in highly accurate variational study \cite{Drake:2014}: for $Z$ slightly smaller that $Z_c$, $Z < Z_c$ normalizable trial functions guarantee convergence in energy up to 14 decimal figures without loosing normalizability!}.
It seems true inside the accuracy of a given variational procedure. It must be noted that even highly accurate, normalizable trial function found at $Z > Z_c$  does not guarantee the existence of the solution of the Schr\"odinger equation in the Hilbert space for $Z < Z_c$ after analytic continuation. Thus, this state is not necessarily bound.

It is worth emphasizing that the expansions (\ref{Zcr}) and
(\ref{ZB}) are exact functionally since they are derived from variational energies emerging
from (\ref{GS}) and (\ref{GSB}) after the exact minimization: minimizing-the-energy
variational parameters are found in a form of expansions near $Z_{c}$ and near $Z_{B}+$, respectively.

\begin{table}
{\footnotesize
\begin{tabular}{|l| l | l | l |}
\hline\hline
\ $Z$   \  &\qquad  $E$ (a.u.) \ &\quad \mbox{fit (ground state)} &\quad \mbox{fit (2nd branch)}\\
\tableline
\ 2.00  \  &\quad  -2.903 724 4    \ &\quad -2.903 719    \ &\quad  -2.201 927   \\
\ 1.00  \  &\quad  -0.527 751 0   \ &\quad  -0.527 749   \ &\quad  -0.515 540   \\
\ 0.94  \  &\quad  -0.449 669 0    \ &\quad  -0.449 668   \ &\quad  -0.447 012   \\
\ 0.92  \  &\quad  -0.425 485 3      \ &\quad  -0.425 485   \ &\quad  -0.424 741   \\
\ $Z^{EBMD}_{c}$ (see (\ref{zcrit})) \ &\quad  -0.414 986 2    \ &\quad -0.414 986   \ &\quad -0.414 793 \\
\hline
\ 0.91  \  &\quad  -0.413 799 2      \ &\quad -0.413 799  \ &\quad -0.413 652   \\
\tableline
\hline
\end{tabular}
}
\caption{
\label{table1S}
 Ground state energy $E$ for a two-electron system at $m_p=\infty$ for selected values of $Z$ found in the
 Lagrange mesh method and rounded to 7 d.d., compared with the fit (\ref{e-fit_1}); the result of the analytic continuation of (\ref{e-fit_1}) to the second branch is shown in the 3rd column.}
\end{table}

It seems natural to check validity of the expansions (\ref{Zcr}) and (\ref{ZB}) for more accurate energies than ones found using (\ref{GS}), (\ref{GSB}). In order to do it we calculate $E(Z)$ in 12 points at $Z \in [0.91 , 1]$\ and\ $Z=2$ in the Lagrange mesh
method accurate to not less than 9 s.d. (see Table \ref{table1S} for illustration).  Then taking into account the points from the domain $Z \in [Z^{EBMD}_{c} , 1]$ {\bf only}, we make a fit using a general terminated Puiseux expansion,
\[
     E\ =\ \sum_{n=0}^{N} a_n (Z - Z_B)^{n/2 + \ep_n} \ ,
\]
for different $N$ assuming that exponents are growing with $n$, searching for optimal values of the parameters $a_n, \ep_n$. As for initial values of parameters, those we use, they are ones taken from the expansion (\ref{ZB}) for (\ref{GS}), (\ref{GSB}). Finally, we arrive at
\[
 E_{1^{1}S}^{(fit)}(Z)\ =\ -\,0.407924\ -\ 1.12347\, (Z - Z_{B})\
 -\ 0.197785\, (Z - Z_{B})^{3/2}\ -\ 0.752842\, (Z - Z_{B})^2
\]
\[
    -\ 0.108259\, (Z - Z_{B})^{5/2}\ -\ 0.014135\, (Z - Z_{B})^3
    +\ 0.00854 \, (Z - Z_{B})^{7/2}
\]
\begin{equation}
\label{e-fit_1}
    +\ 0.00483\, (Z - Z_{B})^4\ -\ 0.000056\, (Z - Z_{B})^{9/2}\ ,
\end{equation}
with the critical charge
\begin{equation}
\label{ZB-1}
    Z_{B}\ =\ 0.904854\ ,
\end{equation}
where as the result of the fit $\ep_n$ turned to be of order $10^{-6} - 10^{-7}$ as well as the parameter $a_1$. It turns out that is artifact of fitting procedure: all these parameters $\ep_n, a_1$ can be placed equal to zero without reducing the quality of the fit!
The expression (\ref{e-fit_1}) reproduces up to a portion of $10^{-6} - 10^{-7}$ in energies at 12 points in $Z$ more or less equally distributed in $Z \in [Z_c, 1.]$ and also the point
$Z=0.91 < Z_c$ (!)\ , for illustration see Table~I (and Table II, see below).
The quality of the fit drops with the increase of $Z$, see Table~I. The numerical coefficients in the expansion decrease with the increase of the order of terms. It may be considered as the indication to the finite radius of convergence of corresponding numerical series of coefficients and thus the Puiseux expansion (\ref{ZB}) itself. We must emphasize that the critical charge $Z_B$ is unphysical: it corresponds to analytically-continued ground state energy to the domain of quasi-stationary states!

In the recent analysis of $1/Z$ expansion \cite{TL:2016} it was shown that the first ten decimal digits in all the first 401 coefficients $e_n$ are found correctly in \cite{Baker:1990}: their weighted sums $E_P(Z)\ =\ \sum^{401} e_n Z^n$ reproduce subsequently the first ten decimal digits in ground state energy of two-electron sequence at $Z=1,2,\ldots 10$\,, found in \cite{Nakashima:2007}. Checking convergence of $1/Z$ expansion we calculate weighted sums $E_P$ for $Z<1$, see Table \ref{E}, and compare them with the result of fit (\ref{e-fit_1}) and with energies obtained in Lagrange mesh calculations on the lattice $92 \times 92 \times 20$. Comparison of $E_P$ and $E_{1^{1}S}^{(fit)}$ indicates to a striking agreement between them: they differ in $\sim 2 \cdot 10^{-7}$ for $Z=0.95$, in $\sim 6 \cdot 10^{-6}$ for $Z=Z^{\infty}_{c} \approx 0.911$ and $\sim 4 \cdot 10^{-5}$ for $Z=0.905$. Hence, the fit (\ref{e-fit_1}) is in agreement with perturbation theory in $1/Z$. In a similar way $E_P$ agrees with
accurate calculations of the ground state energy $E$ in the Lagrange mesh method even for $Z < Z_c$.

\begin{table}[hbt]
\begin{center}
{\small
\begin{tabular}{|l|l|l|l|}
\hline\hline
$Z$		&\qquad $E_P$ \quad &\quad $E_{1^{1}S}^{(fit)}$  (\ref{e-fit_1})  \ &\quad $E$ (Lagr.mesh)\
\\
\tableline					
0.905     \  &\  -0.408 045 \  &\ -0.408 089 \ &\ -0.408 1${}^\dagger$
\\ 						
0.908     \  &\  -0.411 478  \  &\ -0.411 501 \ &\ -0.411 502${}^\dagger$
\\ 						
0.909     \  &\  -0.412 629 \  &\ -0.412 648  \ &\ -0.412 648 5${}^\dagger$
\\ 						
0.91      \  &\  -0.413 783 \  &\ -0.413 799  \ &\ -0.413 799 2
\\[3pt]  					
$Z^{\infty}_{c}$ &\  -0.414 972  \   &\ -0.414 986 \ &\ -0.414 986 2
\\[3pt]           				
0.92      \  &\  -0.425 482 \  &\ -0.425 485  \ &\ -0.425 485 3
\\ 						
0.93      \  &\  -0.437 450  \  &\ -0.437 451  \ &\ -0.437 451 3
\\ 						
0.94      \  &\  -0.449 669 \  &\ -0.449 669  \ &\ -0.449 669 0
\\  						
0.95      \  &\  -0.462 125 \  &\ -0.462 125  \ &\ -0.462 124 7
\\
\tableline \hline
\end{tabular}
}
\end{center}
\caption{\label{E}
 $m_p=\infty$ case: Weighted sum $E_P(Z)$ (a.u.) with $1/Z$ coefficients $e_n$ (first column), see text;
 Energies $E_{1^{1}S}^{(fit)}$  (\ref{e-fit_1}) in a.u. (second column) from the fit based on a terminated Puiseux expansion (\ref{e-fit_1});
 Energies $E$\,(a.u.) in Lagrange mesh method (rounded to 7 s.d., third column), the results marked by ${}^\dagger$ contain confident decimal figures only}
\end{table}

Moreover, it has to be mentioned a remarkable fact that expanding (\ref{e-fit_1}) at $Z=\infty$ the established coefficients in $1/Z$- expansion $e_{10-150}$ are reproduced with relative accuracy $10^{-2}$ (equivalently, the first two significant digits) \cite{TLO:2016}.

\begin{table}
{\footnotesize
\begin{tabular}{|l| l | l | l |}
\hline\hline
\ $Z$        &\qquad       $E$ (a.u.) \ &\quad \mbox{fit (ground state)} \ &\quad \mbox{fit (2nd branch)}\\
\tableline
\ 2.00  \  &\quad  -2.903 304 6 \   & \quad -2.903 178  \ & \quad -2.196 303 \\
\ 1.00  \  &\quad  -0.527 445 9 \ &\quad -0.527 444   \ &\quad -0.515 312 \\
\ 0.96  \  &\quad  -0.474 536 5  \ &\quad  -0.474 537  \ &\quad -0.469 305\\
\ 0.92  \  &\quad  -0.425 242 4  \ &\quad  -0.425 242  \ &\quad -0.424 519\\
\ 0.911069725 (see (\ref{zcrit-mp-us})) \ &\quad  -0.414 798 1 \ &\quad -0.414 798     \ &\quad -0.414 617  \\
\hline
\ 0.91  \  &\quad  -0.413 564 0     \ &\quad -0.413 564       \  &\quad  -0.413 430  \\
\tableline
\hline
\end{tabular}
}
\caption{
\label{table1S-mp}
 Ground state energy $E$ for a two-electron system at $m_p$ for selected values of $Z$, found in the Lagrange mesh method, rounded to 7 d.d. and compared with the fit (\ref{e-fit_2}); the result of analytic continuation of (\ref{e-fit_2}) to the second branch is shown in 3rd column. }
\end{table}
%
%
%
%

Similar analysis to (\ref{e-fit_1}) can be made for the finite-mass case $m_p$. Making a fit for the energies at
$Z \in [Z^{m_p}_{c} , 1]$ (see Table \ref{table1S-mp}) we arrive at
\[
 E_{1^{1}S}^{(fit\,, m_p)}(Z)\ =\ -0.408019 - 1.123511\, (Z - Z_{B}) - 0.198005\, (Z - Z_{B})^{3/2}
   - 0.751842\, (Z - Z_{B})^2\,
\]
\[
    -\, 0.101848\, (Z - Z_{B})^{5/2}\, -\, 0.02171\, (Z - Z_{B})^3 + 0.0039\, (Z - Z_{B})^{7/2}\,
    +\, 0.0126\, (Z - Z_{B})^4\
 \]
\begin{equation}
\label{e-fit_2}
    -\ 0.0028\, (Z - Z_{B})^{9/2}\ ,
\end{equation}
with the critical charge
\begin{equation}
\label{ZB-mp}
    Z_{B}\ =\ 0.90514\ ,
\end{equation}
(cf (\ref{ZB})). Note that the parameters of the fit  (\ref{e-fit_1}) obtained in the static limit change very little  in going from infinite to finite proton mass, see (\ref{e-fit_2}). Both critical charges $Z_{c}, Z_{B}$ increase slightly for the case of the finite proton mass.

It can be immediately seen from Tables I, \ref{table1S-mp} that based on the analytic continuation in $Z$ of the fit (\ref{e-fit_1}), (\ref{e-fit_2}) (it corresponds to the change of sign in front of the terms of half-integer degrees), the existence of the second bound state $2^{1}S$ of negative hydrogen ion is predicted,
\begin{equation}
\label{2S}
    E_{2^{1}S}^{\infty}\ =\ -0.515 54\,a.u.\quad ,
    \quad E_{2^{1}S}^{m_p}\ =\ -0.515 31\,a.u.\ ,
\end{equation}
for the both infinite and finite proton mass, respectively. Transition energies are
\begin{equation}
\label{1S-2S}
    \De E_{1^{1}S \rar 2^{1}S}^{\infty}\ =\ -0.0122\,a.u.\quad ,
    \quad E_{1^{1}S \rar 2^{1}S}^{m_p}\ =\ -0.0121 \,a.u.\ ,
\end{equation}
respectively. H${}^-$ is a "maximally" strongly-correlated Coulomb system, thus,
it is an example of a general 3-body Coulomb problem, it should be treated in full generality. So far, we are unable to find accurately the wavefunction for the $2^{1}S$ state. It is evident it should {\it not} be of the type $(1s2s)$ as for He, Li$^+$ and other two-electron positive ions. Likely, relativistic effects (spin-orbit, spin-spin interactions, radiative effects) to energy of the lowest states (\ref{2S}),(\ref{1S-2S}) should not be more than $\sim 10^{-4}$\,a.u. similarly to ones for Helium \cite{Alexander:2010}

Note that conceptually the prediction (\ref{2S}) is in agreement with D.R.~Yafaev's rigorous
mathematical statement \cite{Yafaev:1972} about the finite number of bound states at $Z=1$ but in
contradiction to the widely-known theoretical statement by R.N.~Hill \cite{Hill,Hill2} about the
existence of a single, spin-singlet bound state of H${}^-$ only. Mathematical justification of the
latter result is absent (to the best of the present author's knowledge). It seems it is a challenge
to check the prediction (\ref{2S}) experimentally since the transition $1^{1}S \rar 2^{1}S$ is the
forbidden type transition, hence, it should be a type of multi-photon transition and/or
bound-free-bound transitions, which is difficult to observe. In principle, the spectra of H${}^-$
contains one more, but metastable spin-triplet $2^{3}P$ state \cite{Drake:1970}.  
Note that for helium atom, the main modes are (permitted) electric-dipole transitions 
$2^{1}S \rar 2^{1}P$ and $2^{1}P \rar 1^{1}S$, see \cite{Dalgarno:1966, VanDyck:1971}.  
Experimental study of the (forbidden) transition $1^{1}S \rar 2^{1}S$ for helium atom 
was done only recently using the Doppler-free two-photon spectroscopy, see \cite{Eyler}.

In present paper we studied the domain of bound states of three-body Coulomb problem $Z > Z_c$ and the analytic continuation to the complex $Z$-plane of the ground state $1^{1}S$, in particular, to $Z < Z_c$. We showed that the second critical charge predicted in \cite{Stillinger:1966} is unphysical: 
it appears in analytic continuation of the ground state energy to $Z < Z_c$, thus, it can not 
be measured. It seems important to carry out a similar study for the domain of quasi-stationary 
states $Z < Z_c$, in particular, checking the analytic discontinuity $Z = Z_c$.

AVT thanks G.~W.~F.~Drake (Windsor), N.~Berrah, E.~Eyler, W.~Smith and W.~Stwalley (Storrs), J.-P.~Karr (Paris) for the interest to the work and useful discussions. One of us (AVT) is grateful to Stony Brook University for the kind hospitality during his sabbatical stay, where a part of the research was done, he was supported in part by PASPA program (Mexico). The final stage of the present work was carried out at Simons Center for Geometry and Physics (Stony Brook).
The research by JCLV and AVT is supported in part by DGAPA grant IN108815 and CONACyT grant 166189 (Mexico).


\end{document}